\documentclass[12pt]{article}
\usepackage{amsfonts}

\newtheorem{theorem}{Theorem}
\newtheorem{lemma}{Lemma}
\newtheorem{proposition}{Proposition}
\newtheorem{corollary}{Corollary}
\newtheorem{definition}{Definition}
\def\Proof{\noindent{{\bf Proof.}\ }}
\def\QED{{\hfill \frame{\phantom{I}}}\\}

\def\supp{\mathop{\rm supp}}
\def\<{{\langle}}
\def\>{{\rangle}}
\def\Prob{\mathop{\rm Prob}}
\def\Meas{{\rm Meas}}
\def\Borel{{\cal B}}
\def\Real{{\Bbb{R}}}
\def\Zahl{{\mathbb{Z}}}
\def\Ntr{{\mathbb{N}}}
\def\Remark{\noindent{\bf Remark.}\ }

\begin{document}
\title{ Uniqueness of the EPR--chemeleon model  }
\author{
Luigi Accardi\thanks{Email: accardi@volterra.uniroma2.it}\\
Centro Vito Volterra,\\
 Universit\`a degli Studi di Roma
``Tor Vergata", \\
Via Columbia, Rome 00133, Italy\\
Satoshi Uchiyama\thanks{Email: uchiyama@hokusei.ac.jp}\\
Department of Life and Creative Sciences \\
Hokusei Gakuen University Junior College\\
Atsubetsu-ku, Sapporo 004-8631, Japan\\
}


\maketitle

\begin{abstract}
A classical deterministic, reversible dynamical systems, reproducing 
the Einstein--Podolsky--Rosen (EPR) correlations in full respect of causality and locality 
and without the introduction of any ad hoc selection procedure, was constructed in 
the paper \cite{AIR02}.

In the present paper we prove that the above mentioned model is unique 
(see Theorem (\ref{ModulusCos}) ) in the sense that any local
causal probability measure which reproduces the EPR correlations must coincide, 
under natural and generic assumptions, with the one constructed in \cite{AIR02}. 
\end{abstract}

\maketitle


\section{Introduction}

It is now understood:
\begin{itemize}
\item[(i)]
that the the common mathematical root of the apparent paradoxes arising in 
connection with $2$--slit type or EPR type experiments is that certain statistical data (conditional probabilities, correlations, $\dots$) cannot be reproduced  by a single Kolmogorovian probability space \cite{[Ac81a]} 

\item[(ii)]
that there exist classical deterministic, reversible dynamical systems, reproducing 
the singlet correlations of spins pairs (or of polarizations of a pair of 
entangled photons), called EPR correlations in the following, \cite{AIR02, AIR03}.
\end{itemize}
The construction of such dynamical systems was made possible by a new physical idea 
(the chameleon effect) and a new mathematical tool (the notion of notrivial local causal
measure). 

The chameleon effect consists in the statement that the local dynamics of some systems 
(adaptive systems) may depend on the observable that one measures.
The purpose of the EPR-chameleon model is a simple realization of this general idea.

The striking feature of the EPR-chameleon model is that the dynamics of each spin as well 
as the structure of the state (i.e. the probability measure defining the statisitics)
is local and causal i.e., there is no action at distance between the spins in the pair
or between the two measurement apparata and no previous knowledge of the future measurements.
Everything is completely pre--determined at the source through an {\it if--then} scheme
which is typical of adaptive systems and which justifies the chameleon metaphora ({\it if
I meet a leaf I will become green, if I meet a piece of wood I will become brown}).
In the mathematical model the {\it if--then} scheme is entirely coded in an intrinsic
dynamics and an initial state and no artificial selection or rejection procedures are introduced 
by hands.

Even if the models described in the present paper are inspired to the EPR--Bohm type experiments
\cite{EPR, Bohm}, we emphasize that all our constructions will be entirely within 
the classical theory of dynamical systems.

The organization of the paper is as follows: 

Section 2 introduces the notion of triviality of a LC measure and shows that such measures
cannot violate Bell's inequality. Thus if we want to reproduce the EPR--type correlations,
then we must investigate nontrivial LC measures.

The main result of Section 2 is the proof of the fact that the class of trivial LC 
measures and the class of nontrivial LC measures cannot be connected by any local 
and reversible dynamics (Corollary \ref{AccThrm}).

Section 3 contains the main result of the present paper i.e. the proof (see Theorem 
(\ref{ModulusCos}) ) of the fact that any LC probability measure which reproduces the EPR correlations must coincide, under natural and generic assumptions, with the one proposed in \cite{AIR02}.

Section 4 makes explicit the mathematical differences between passive and adaptive dynamical
systems (see also (\cite{AIR03})).

Section 5 shows how the difference between standard and distant particles empirical 
correlations is reflected in the corresponding mathematical models.

The generic assumptions used in the proof of our uniqueness theorem  (Theorem (\ref{ModulusCos}) ) 
are the following:
\begin{itemize}
\item[(i)]   The condition of statistical pre--determination (see Definition (\ref{statpred}))

\item[(ii)]  The rotation invariance of the densities describing the local apparata (see condition 
(\ref{rotinvcond}))

\item[(iii)]  The twice continuous differentiability of these densities (see Theorem (\ref{ModulusCos}))

\item[(iv)]  The absolute continuity of the source measure with respect to the Lebesgue measure
(see Proposition (\ref{DiagonalSupportZero})).
\end{itemize}
While conditions (i) and (ii) have a natural physical interpretation, we don't see ant natural 
physical justification for conditions (iii) and (iv). 

For example at the moment we have no reasons to exclude the possibility of reproducing
the EPR correlations with a source measure having a fractal support.

Therefore it would be interesting to know  if, by dropping some of these assumptions, 
the uniqueness result continues to be true. This problem will be the object of further investigations.


\section{Trivial LC measures}\label{triv-lc-meas}

We consider a composite system made up of two subsystems, often called ``particles" and denoted with the symbos $1$ and $2$ respectively. Their ``configuration" (or ``phase") spaces will be denoted by
$S_{1}$ and  $S_{2}$ respectively. 
The two systems are spatially separated so that the mutual interactions between them can be neglected. Each system interacts locally with a measurement apparatus, i.e. system $1$ with 
apparatus $m_{1}$ and system $2$ with apparatus $m_{2}$. The configuration spaces of the measurement apparata  will be denoted by $M_{1}$ and $M_{2}$ respectively. 
We use the indices $a,b, \dots \in I$ to represent settings of the measurement apparata.
In the second part of the paper
from section (\ref{airmod}) on we specialize the set of indices $I$ to be the interval $[0,2\pi]$.

The notion of ``local and causal probability measure" is crucial for EPR-chameleon models.
\begin{definition} (\cite{AIR03}, Definition 6.)
A probability measure $P_{a, b}$ on \hfill\break
$S_{1}\times S_{2} \times M_{1} \times M_{2}$ is called  
local and causal (LC, shortly) if it has the form
\begin{equation}\label{LCmeasure}
dP_{a, b}(s_{1}, s_{2}, \lambda_{1}, \lambda_{2} ) = dP_{S}(s_{1}, s_{2}) P_{1, a}(d\lambda_{1}; s_{1}) P_{2, b}(d\lambda_{2}; s_{2}),
\end{equation}
where $P_{S}$ is a probability measure on $S_{1}\times S_{2}$;
for all $s_{1} \in S_{1}$, $P_{1, a}(\ \cdot\ ; s_{1})$ is a positive measure on $M_{1}$;
for all $s_{2} \in S_{2}$, $P_{2, b}(\ \cdot\ ; s_{2})$ is a positive measure on $M_{2}$.
\end{definition}

Notice that the requirement that $P_{S}$ is a probability measure on $S_{1}\times S_{2}$
is not essential: if $P_{S}$ is any finite measure, by multiplying $P_{S}$, 
$P_{1, a}(\ \cdot\ ; s_{1})$ and $P_{2, b}(\ \cdot\ ; s_{2})$ by positive constants
whose product is equal to $1$, one can always reduce oneself to the case that $P_{S}$ 
is a probability measure.

This multiplication and division by the same constant is trivial from the mathematical 
point of view, but it may be essential for the purpose of a local simulation of a LC
measure (see the discussion in section (\ref{empcordistpa}) below). 
This is precisely the case for the measure constructed in \cite{AIR03}.

Let us assume that all the followings are compact Hausdorff spaces:
\begin{itemize}
\item[--] the configuration space $S_{1}$ of the subsystem $1$, 

\item[--] the configuration space $S_{2}$ of the subsystem $2$, 

\item[--] the configuration space $M_{1}$ of the measurement apparatus for the subsystem $1$, 

\item[--] the configuration space $M_{2}$ of the measurement apparatus for the subsystem $2$. 
\end{itemize}
In terms of these we define the configuration spaces for the composite systems:
$$
S:= S_{1}\times S_{2} \ ; \ M:=M_{1} \times M_{2}\ ; \ 
\Omega_{1} := S_{1}\times M_{1} \ ; \ \Omega_{2} := S_{2}\times M_{2}
$$
\begin{equation}\label{df-omega}
\Omega := \Omega_{1} \times \Omega_{2} = S_{1}\times M_{1}\times S_{2}\times M_{2}
 = S_{1}\times S_{2}\times M_{1}\times M_{2}.
\end{equation}
Let $\Meas(\Omega)$ denote the set of all regular, signed, finite Borel measures on $(\Omega, \Borel)$.
$\<\Meas(\Omega),\ C(\Omega)\>$ denotes the duality $\Meas(\Omega)= C(\Omega)^{*}$.
$\Meas_{+}(\Omega)$ and $\Prob(\Omega)$ denote  the set of all positive measures and  the set of all probability measures in $\Meas(\Omega)$ respectively.

Then, since $P_{S}$ is a probability measure on $ S_{1}\times S_{2}$,  
$P_{a, b}$, given by (\ref{LCmeasure}), is a LC measure on 
$S_{1}\times S_{2} \times M_{1} \times M_{2}$ which can be written in the following functional form:
\begin{equation}
P_{a, b} := 
P_{S} \circ  (\overline{P}_{1,a}\otimes \overline{P}_{2,b}) \in ( C(\Omega_{1}) \otimes C(\Omega_{2}) )^{*} 
= C(\Omega_{1}\times \Omega_{2})^{*}, 
\label{FunctionalForm}
\end{equation}
where, for $j=1, 2$ and $x= a, b$, the linear maps
$$
\overline{P}_{j, x}: C(\Omega_{j})=C(S_{j}\times M_{j})\to   C(S_{j})\subseteq C(\Omega_{j})
$$ 
are defined by
\begin{equation}
\overline{P}_{j,x}(f)(s_{j}) := \int_{M_{j}} f(s_{j}, \lambda_{j}) dP_{j,x}(\lambda_{j} ;s_{j})
\label{IntegMj}
\end{equation}
for each  $f\in C(S_{j}\times M_{j})$.

\begin{definition}\label{dftriv-pm} 
(\cite{AIR03}, Definition 7.)
A LC probability measure on the space $S_{1}\times S_{2} \times M_{1} \times M_{2}$ 
\[
dP_{a, b}(s_{1}, s_{2}, \lambda_{1}, \lambda_{2} ) = dP_{S}(s_{1}, s_{2}) dP_{1, a}(\lambda_{1}; s_{1}) dP_{2, b}(\lambda_{2}; s_{2})
\]
is called trivial if, in the notation (\ref{IntegMj}), $\forall a,b\in I$ the map 
$$
\overline{P}_{1,a}\otimes \overline{P}_{2,b} \ : \ C(\Omega_{1}\times \Omega_{2})\to C(S_{1}\times S_{2})  
$$
is a $P_{S}$--conditional expectation i.e.
\begin{equation}\label{pjbarce}
\overline{P}_{1,a}(1_{1})(s_{1}) \overline{P}_{2,b}(1_{2})(s_{2}) \equiv 1
\qquad , \qquad P_{S}\mbox{-a.e.}
\end{equation}
\end{definition}

Denoting
\begin{equation}\label{p1as1}
p_{1, a}(s_{1}) := \overline{P}_{1,a}(1_{1})(s_{1})=\int_{M_{1}} dP_{1, a}(\lambda_{1}; s_{1})
\end{equation}
\begin{equation}\label{p2bs2}
p_{2, b}(s_{2}) :=\overline{P}_{2,b}(1_{2})(s_{2})= \int_{M_{2}} dP_{2, b}(\lambda_{2}; s_{2})
\end{equation}
condition (\ref{pjbarce}) becomes equivalent to:
\begin{equation}
p_{1, a}(s_{1}) p_{2, b}(s_{2}) = 1  \qquad ,\qquad \ P_{S}\mbox{-a.e.}
\end{equation}

\Remark
If a LC measure is trivial, then from
\[
p_{1, a}(s_{1}) = \frac{1}{ p_{2, b}(s_{2})}\qquad ,\ P_{S}\mbox{-a.e.},
\]
there exists a positive real number $c$ such that
\[
p_{1, a}(s_{1}) = c,\
p_{2, b}(s_{2}) = \frac{1}{c}, \ P_{S}\mbox{-a.e.}
\]
By redefining $P_{1, a}':= (1/c)P_{1, a}$ , $P_{2, b}' :=c P_{2, b}$, we can assume without loss of generality that
\[
p_{1, a}(s_{1}) = 1,\ 
p_{2, b}(s_{2}) = 1 \qquad  , \ P_{S}\mbox{-a.e.}
\]
\medskip

The following result shows why contextuality alone is not sufficient to account for the
violation of Bell's inequality.

\begin{proposition} (\cite{AIR02})
 Let $I$ be any index set and let $P_{a, b}$ ($a,b\in I$) be a family of 
trivial LC probability measures on the space $\Omega$ defined by (\ref{df-omega}).
Then the pair correlations of any family of random variables 
$S_{a}^{(1)}, S_{b}^{(2)} : \Omega\to [-1,1]$ ($a,b\in I$) satisfying the locality condition
$$
S_{a}^{(1)}(\omega_1, \omega_2)=S_{a}^{(1)}(\omega_1) \ ; \
S_{b}^{(2)}(\omega_1, \omega_2)=S_{b}^{(2)}(\omega_2) \ ; \
(\omega_1, \omega_2)\in \Omega = \Omega_1\times \Omega_2
$$
cannot violate Bell's inequality.
\end{proposition}
\Proof
The pair correlations of the random variables $S_{a}^{(1)}, S_{b}^{(2)}$ are defined by
\[
C(a, b):= \< P_{a, b},\ S_{a}^{(1)}\otimes S_{b}^{(2)} \>
=\< P_{S},\ \overline{P}_{1,a}( S_{a}^{(1)} )\otimes \overline{P}_{2, b}(S_{b}^{(2)}) \>_{S_{1}\times S_{2}}.
\]
Using the functional form (\ref{FunctionalForm}) of the trivial measures $P_{a, b}$ one finds
\begin{eqnarray*}
&&|C(a, b) -C(a,b')| + |C(a', b) + C(a', b')| \\
&\leq&
\<P_{S},\  | \overline{P}_{1,a}( S_{a}^{(1)} )\otimes [\overline{P}_{2, b}(S_{b}^{(2)}) -
\overline{P}_{2, b'}(S_{b'}^{(2)})] | \>_{S_{1}\times S_{2}}\\
&&+
\<P_{S},\  | \overline{P}_{1,a'}( S_{a'}^{(1)} )\otimes [\overline{P}_{2, b}(S_{b}^{(2)}) +
 \overline{P}_{2, b'}(S_{b'}^{(2)})]| \>_{S_{1}\times S_{2}} \\
&\leq&
\<P_{S},\  | \overline{P}_{2, b}(S_{b}^{(2)}) -
\overline{P}_{2, b'}(S_{b'}^{(2)}) | +  |\overline{P}_{2, b}(S_{b}^{(2)}) +
 \overline{P}_{2, b'}(S_{b'}^{(2)}) | \>_{S_{1}\times S_{2}} \leq 2
\end{eqnarray*}
where in the last inequality we have used the fact that Bell's inequality (in CHSH form)
is satisfied by any quadruple of random variables, on a single probability space, with values
in the interval $[-1,1]$ (for a proof of this statement see \cite{AR00, Bell71}
\QED

\Remark
To be a trivial LC measure is a sufficient, but not necessary contition to satisfy Bell's inequality.
There are nontrivial LC measures which are essentially trivial and do not violate Bell's inequality.
For example let $P_{a, b} = P_{S}\circ ( \overline{P}_{1, a}\otimes \overline{P}_{2, b} )$ be a trivial LC measure.
Let $q_{1}$ and $q_{2}$ be non-zero measurable functions on $S_{1}$ and $S_{2}$ respectively such that
\[
\int_{S_{1}\times S_{2}} \!\!\!\! dP_{S}(s_{1}, s_{2}) \ q_{1}(s_{1}) q_{2}(s_{2}) = 1.
\]
Define $Q \in \Prob(S_{1}\times S_{2})$ by $dQ(s_{1}, s_{2}):= q_{1}(s_{1}) q_{2}(s_{2}) dP_{S}(s_{1}, s_{2})$.
Since
\[
P_{a, b} = Q\circ \left( \left( (1/q_{1})\overline{P}_{1, a} \right) \otimes \left((1/q_{2})\overline{P}_{2, b}\right) \right),
\]
if $q_{1}\otimes q_{2}$ is not constant on $\supp Q$, then $P_{a, b}$ becomes nontrivial.
\medskip

Recall that, for any pair of compact topological spaces $\Omega$, $S$, a linear map
$${\cal T}^*: C(\Omega)\to  C(S)$$
is called a Markov operator if it is positivity preserving ($f\geq0\Rightarrow{\cal T}^*(f)\geq0$, $f\in{\cal C}(\Omega)$) and 
$${\cal T}^*(1_\Omega)=1_S$$
If on $S$ there is a probability measure $P_S$ and ${\cal T}^*$ satisfies the weaker conditions
$$f\geq0\Rightarrow{\cal T}^*(f)\geq0\ ;\quad P_S\hbox{-a.e. }f\in{\cal C}(\Omega)$$
$${\cal T}^*(1_\Omega)=1_S\ ,\quad P_S\hbox{-a.e.}$$
we call it a $P_S$--Markov operator. Now let
$$\Omega=\Omega_1\times\Omega_2\ ;\quad S=S_1\times S_2.$$
The identifications:
$$s_1\equiv s_1\times S_2\ ;\quad s_2\equiv S_1\times s_2\ ;\quad s_1\in S_1,\ s_2\in S_2$$
allows us to consider both $S_1$ and $S_2$ as subsets of $S_1\times S_2$.

\begin{lemma}\label{eqcondmark} 
For $j=1,2$, let ${\cal T}^*_j: C(\Omega_j)\to C(\Omega_j)$ be a positivity preserving linear operator.
The following conditions are equivalent:
\begin{equation}\label{t-triv}
\overline P_{1,a}({\cal T}^*_1(1))\overline P_{1,b}({\cal T}^*_2(1))=1\ ;\ P_S-\hbox{a.e.}
\end{equation}
there exists a constant $c>0$ such that
\begin{equation}\label{cep}
\overline P_{1,a}(c{\cal T}^*_1(1))=\overline P_{1,b}\left({\cal T}^*_2(1)/c\right)=1\ ;
\ P_S\hbox{-a.e.}
\end{equation}
\end{lemma}

\noindent{\it Proof\/}. It is clear that (\ref{cep}) $\Rightarrow$ (\ref{t-triv}). 
Let us prove the converse implication. If (\ref{t-triv}) holds, then
$$
P_S\circ([\overline P_{1,a}\circ{\cal T}^*_1]\otimes[\overline P_{2,b}\circ{\cal T}^*_2])
$$
is a trivial measure.
Therefore, by the remark after Definition (2) there exists a constant $c>0$ such that
$$
c\overline P_{1,a}({\cal T}^*_1(1))(s_1)={1\over c}\,\overline P_{2,b}({\cal T}^*_2(1))(s_2)=1\qquad 
; \quad P_S-\forall\,(s_1,s_2)\in S_1\times S_2
$$
and this is (\ref{cep}).

\begin{definition}\label{pab-mark}
A linear positive operator 
${\cal T}^*_1\otimes{\cal T}^*_2 : C(\Omega_1\times\Omega_2)\to C(\Omega_1\times\Omega_2)$ 
(or equivalently its dual ${\cal T}_{1}\otimes{\cal T}_{2}$, acting on measures), which satisfies the conditions of Lemma 
(\ref{eqcondmark})  will be called a $P_{a,b}$--Markovian operator.
In such a case, by absorbing the constants $c$, $1/c$ in the definition of ${\cal T}^*_1$ and
${\cal T}^*_2 $, one can always assume that they are equal to $1$.
\end{definition}

{\bf Remark}. Notice that any Markovian operator is  $P_{a,b}$--Markovian for any $P_{a,b}$.


\begin{theorem} \label{Trvl2TrvlP0}
Let, for $j = 1, 2$, ${\cal T}_{j}$ be a linear mapping of 
$\Meas_{+}(\Omega_{j})$ into $\Meas_{+}(\Omega_{j})$  
such that ${\cal T}_{j}^{*}: C(\Omega_{j}) \to C(\Omega_{j})$ and let
$$
P_{a, b}= P_{S}\circ(\overline{P}_{1,a}\otimes \overline{P}_{2, b}) \in \Prob(\Omega_{1}\times \Omega_{2})
$$
be any trivial LC measure. Then if ${\cal T}_{1, a}\otimes{\cal T}_{2, b}$ is a $P_{a, b}$--Markovian 
operator, \hfil\break
$({\cal T}_{1, a}\otimes {\cal T}_{2, b})(P_{a, b})$ is a trivial LC measure.
In particular, if ${\cal T}_{1, a}\otimes{\cal T}_{2, b}$ is a Markov operator, it maps
trivial LC measures into trivial LC measures.
\end{theorem}
\Proof
The functional form of $({\cal T}_{1, a}\otimes {\cal T}_{2, b})(P_{a, b})$ is:
\begin{equation}
({\cal T}_{1, a}\otimes {\cal T}_{2, b})(P_{a, b}) = P_{S} \circ  
(\overline{P}_{1,a}\circ {\cal T}_{1, a}^{*}\otimes \overline{P}_{2,b}\circ  {\cal T}_{2, b}^{*}).
\end{equation}
Condition (\ref{cep}) (with $c=1$) is equivalent to
$$
\overline P_{1,a}({\cal T}^*_1(1))=\overline P_{2,b}\left({\cal T}^*_2(1)\right)=1\ ;
\ P_S\hbox{-a.e.}
$$
which is equivalent to the triviality of $({\cal T}_{1, a}\otimes {\cal T}_{2, b})(P_{a, b})$.

\begin{corollary} \label{AccThrm}
Any local reversible dynamics induces a mapping which maps a nontrivial (resp. trivial) LC measure 
into a nontrivial (resp. trivial) LC measure.
\end{corollary}
\Proof
The statement about trivial LC measures follows from Theorem (\ref{Trvl2TrvlP0}). 

Let $\mu$ be a nontrivial LC measure and $T$ be a reversible measurable transformation of 
$S_{1}\times M_{1} \times S_{2}\times M_{2}$ into itself.
Suppose by contradiction that $\nu := \mu\circ T$ is trivial.
 
The linear mapping ${\cal T}$ induced by $T$ is a Markov operator satisfying
$\mu = {\cal T}(\nu) := \nu \circ T^{-1}$. Its inverse is also a Markov operator satisfying
$\nu = {\cal T}^{-1}(\mu ) := \mu  \circ T$. 

But if $T$ is local i.e. of the form $T = T_{1}\times T_{2}$ for some
$T_{1}:S_{1}\times M_{1}\to S_{1}\times M_{1}$ and $T_{2}:S_{2}\times M_{2}\to S_{2}\times M_{2}$ 
, then ${\cal T}={\cal T}_1\otimes{\cal T}_2$ where ${\cal T}_1$ and ${\cal T}_2$ are Markov operators.
By the remark after Definition (\ref{pab-mark}) this contradicts Theorem \ref{Trvl2TrvlP0}.
\QED


\section{AIR models}\label{airmod}

In the EPR-chameleon model constructed in \cite{AIR02, AIR03} (hereinafter AIR model),  
which reproduces the EPR--Bohm correlations, the configuration space of the single particle 
is chosen to be the unit circle, i.e.
$$S_1=S_2=S^1:=\{(x,y)\in\Bbb R^2:x^2+y^2=1\}$$
and the observables to be functions $f:S^1\to\Bbb R$. It is convenient, in order to calculate 
easily the integrals expressing the correlations, to identify $S^1$ with the quotient space 
$\Bbb R/(2\pi\Zahl)\equiv [0,2\pi)$, i.e. the real numbers defined modulo $2\pi$ 
and the observables with periodic functions $f:\Bbb R\to\Bbb R$ with period $2\pi$.
We will freely use this identification in the following.
$S_{1}\times S_{2}$ is a two-dimensional torus  $T^{2} := S^{1}\times S^{1}$.
Define
\[
I_{a} := \left[- \frac{\pi}{2} + a,\  a + \frac{\pi}{2} \right),
\]
\[
J_{a} := \left[a +\frac{\pi}{2},\  a + \frac{3\pi}{2} \right) 
\]
Under our convenction of identifying numbers modulo $2\pi$, one has 
$$
I_{a + \pi} = J_{a} \qquad , \qquad J_{a+\pi} = I_{a}
$$

The random variables  $S_{a}^{(1)}$ and $S_{b}^{(2)}$, representing outcomes of 
measurements of spins, are parametrized by $a, b \in [0, 2\pi) $ and are defined by
\begin{eqnarray}
S_{a}^{(1)}(s_{1}) &:=& \chi_{I_{a}}(s_{1}) - \chi_{J_{a}}(s_{1}) \qquad , \ s_{1}\in S_{1}\\
S_{b}^{(2)}(s_{2}) &:=& -\chi_{I_{b}}(s_{2}) + \chi_{J_{b}}(s_{2})  \qquad , \ s_{2}\in S_{2}
\end{eqnarray}
thus they depend only on the final configurations of the particles, $s_{1}\in S_{1}$ and 
$ s_{2}\in S_{2}$ respectively and are independent of the (final) configurations of the measurement apparata
(the reason why we interpret these points as final rather than as initial configurations is discussed in
sections (\ref{2setfordet}) and (\ref{empcordistpa})).

In the present section we study the most general family of local causal probability measures on 
$T^{2}$ which reproduce the EPR--Bohm correlations and we prove that, under natural generic conditions, they must have the form used in the  AIR model.

If $P_{a, b}$ is a local causal probability measure on $S_{1}\times S_{2} \times M_{1} \times M_{2}$ 
of the form (\ref{LCmeasure}), we denote $R_{a,b}$ its marginal probability on $T^{2} = S^{1}\times S^{1}$. 
Using the notations (\ref{p1as1}), (\ref{p2bs2}), we can write $R_{a,b}$ in the following form:
\begin{equation}
d R_{a, b}(s_{1}, s_{2}) = d P_{S}(s_{1}, s_{2}) \ p_{1,a}(s_{1}) p_{2, b}(s_{2}),
\label{ChamMeas}
\end{equation}
where $s_{1}, s_{2} \in [0, 2\pi)$ are fixed parameterizations of $S_{1} = S^{1}$ and 
$S_{2}=S^{1}$ respectively, $P_{S}$ is a probability measure on $T^{2}$ and 
$p_{1,a}(s_{1})$, $p_{2,b}(s_{2})$ $\geq 0$. 

We say that the family of probability measures (\ref{ChamMeas}) reproduces the statistics of the EPR--Bohm experiment if, for any $a,b \in [0, 2\pi)$ one has:
\begin{eqnarray} \label{QMP}
R_{a, b}( I_{a}\times I_{b})
&=& \frac{1}{2}\cos^{2}\left(\frac{b -a }{2}\right)
=: P_{a,b}^{+-}\\ \nonumber
R_{a, b}(J_{a} \times J_{b})
&=& \frac{1}{2}\cos^{2}\left(\frac{b -a }{2}\right)
=: P_{a, b}^{-+}\\ \nonumber
R_{a, b}( I_{a} \times J_{b})
&=& \frac{1}{2}\sin^{2}\left(\frac{b -a }{2}\right)
=: P_{a, b}^{++}\\ \nonumber
R_{a, b}( J_{a}\times I_{b})
&=& \frac{1}{2}\sin^{2}\left(\frac{b -a }{2}\right)
=: P_{a, b}^{--}.   \nonumber
\end{eqnarray}

\Remark
Let us fix (arbitrarily) a single oriented reference framework for the whole experiment, 
determined by 3 orthogonal axes $x$, $y$, $z$.
We assume that the trajectories of all particles entirely lay in the $(x,y)$--plane and 
that the parameters $a$ and $b$ represent the angles of the orientation of the spin 
analyzers with the $x$--axis.

The identities $(\ref{QMP})$ show that the experimental probabilities do not depend on the 
arbitrarily chosen global reference frame but, as one would expect intuitively, only on 
the relative orientation of the spin analyzers. Given our assumptions,
this invariance of $(\ref{QMP})$ expresses the invariance of the experimental
probabilities under rotations around the $z$-axis, i.e. under transformations of the form 
$a \mapsto a + c$ and $b \mapsto b + c$ for any real number $c$:
$P_{a, b}^{++} = P_{a + c, b + c}^{++}$, etc.
Choosing $c = - a$ or $-b$ , this implies that $P_{a, b}^{++} = P_{a - b, 0}^{++}= P_{0, b-a}^{++}$, etc.
This suggests the following:

\begin{definition}\label{empeqpm} 
Two probability measures $R_{a,b}$, $R_{a',b'}$, of the family (16), are called 
empirically equivalent if thy reproduce exactly the same empirical data, i.e. if:
$$R_{a,b}(I_a\times I_b)=R_{a',b'}(I_{a'}\times I_{b'})$$
$$R_{a,b}(J_a\times I_b)=R_{a',b'}(J_{a'}\times I_{b'})$$
$$R_{a,b}(I_a\times J_b)=R_{a',b'}(I_{a'}\times J_{b'})$$
$$R_{a,b}(J_a\times J_b)=R_{a',b'}(J_{a'}\times J_{b'})$$
\end{definition}

Denoting $\sim$ the relation of empirical equivalence among probability measures and using the terminology of Definition (\ref{empeqpm}), the rotation invariance property of the family of probability measures (\ref{ChamMeas}), can be reformulated as follows:
\begin{equation}
R_{a,b}\sim R_{a-b,0}\sim R_{0,b-a}\ ;\quad\forall\,a,b\in[0,2\pi)\label{rotinvr}.
\end{equation}
\medskip
Notice however that the rotation invariance of the experimentally measured probabilities 
is a weaker condition than the rotation invariance of the full probability measures.

\subsection{ The support of $R_{a, b}$}

Let us consider a measurable space $(\Omega, \Borel)$ consisting of a compact Hausdorff space 
$\Omega$ and its Borel $\sigma$-algebra $\Borel$  generated by the open sets of $\Omega$.

\begin{definition}
For $P \in \Prob(\Omega)$ (the set of all probability measures on $\Omega$), put
${\cal F} := \{ A \in \Borel \ : \ A \mbox{ is open and } \ P(A)=0 \}$
and define 
$\supp P := \left( \bigcup_{A \in {\cal F}} A \right)^{c}$.
We call $\supp P$ the support of $P$.
\end{definition}

Define the diagonal subset  $\Delta$ of $T^{2}$  by
\begin{equation}
\Delta := \left\{(s_{1}, s_{2} )\in T^{2} :\ s_{1} = s_{2}  (\mbox{mod } 2\pi) \right\}.
\end{equation} 

\begin{definition}\label{statpred} 
The family (\ref{ChamMeas}) of probability measures satisfies the condition of 
{\bf statistical pre--determination\/} if $\forall (s_{1}, s_{2}) \in T^{2} \setminus \Delta$ 
there exists $a \in S^1$ and a neighborhood $G$ of $(s_{1}, s_{2})$, contained in 
$(I_{a}\times J_{a}) \cup (J_{a}\times I_{a})$ such that 
$$
p_{1,a}(s_{1}') p_{2, a}(s_{2}') > 0 \qquad ; \qquad \forall (s_{1}', s_{2}') \in G.
$$
\end{definition}
\noindent{\bf Remark}. 
If $S_1=S_2$ were a discrete space, the condition $R_{a,a}(s_1,s_2)=0$ would define 
the forbidden configurations for the pair of observables $S_{a}^{(1)}(s_{1}) $,
$S_{a}^{(2)}(s_{2}) $, i.e. those configurations which give zero contribution to the
correlation of these observables.

Statistical predetermination means that, the fact that a configuration is statistically 
forbidden for such all measurements that the outcomes are precisely (anti-) correlated cannot depend on the local measurements, but it is 
defined at the source. 

Since our configuration space is not discrete, we introduce the neighborhood $G$, 
of $(s_{1}, s_{2})$, to express this idea.

\begin{proposition} \label{DiagonalSupportZero}
Suppose that the family of probability measures $(\ref{ChamMeas})$ satisfies (\ref{QMP}) 
(agreement with the empirical data) and the condition of statistical pre--determination.
Then 
$$
supp P_{S} \subseteq \Delta
$$
In particular, if the restriction of $P_{S}$ to $\Delta$ is absolutely continuous with 
respect to the Lebesgue measure on $\Delta$, then there exists a nonnegative function 
$\rho(s_{1})$ on $\Delta\equiv S^1$ such that: 
\begin{equation}\label{PSconcd}
dP_{S}(s_{1}, s_{2}) = \rho(s_{1}) \delta(s_{1} - s_{2}) ds_{1} ds_{2}
\end{equation} 
\end{proposition}
\Proof
By assumption, for each $(s_{1}, s_{2})\in T^{2} \setminus \Delta$, there exist $a \in [0, 2\pi)$ 
and a neighborhood $G$ of $(s_{1}, s_{2})$ contained in 
$(I_{a}\times J_{a})\cup(J_{a}\times I_{a}) $ such that
\begin{eqnarray*}
\int_{S_{1}\times S_{2}}   d P_{S}\ p_{1, a}\otimes p_{2, a} \cdot \chi_{G} &=&R_{a,a}(G) \leq R_{a, a}((I_{a}\times J_{a}) \cup (J_{a}\times I_{a}))\\
 &=& P_{a,a}^{++} + P_{a,a}^{--} = 0.
\end{eqnarray*}
Since  $p_{1,a}\otimes p_{2, a} > 0$ on $G$, it follows that $P_{S}(G)=0$, i.e.
$G \subseteq (\supp P_{S})^{c}$. Thus any point in $T^{2} \setminus \Delta$ has a neighborhood
contained in $(\supp P_{S})^{c}$. This means that $T^{2} \setminus \Delta \subseteq (\supp P_{S})^{c}$
or equivalently that $\supp P_{S} \subseteq \Delta$.

In view of this property, the existence of $\rho$ is equivalent to the absolute continuity of 
the restriction of $P_{S}$ on $\Delta$.
\QED

\begin{theorem} \label{ModulusCos}
Under the assumptions of Proposition (\ref{DiagonalSupportZero}), if $p_{1, a}$ and 
$p_{2, b}$ are rotation invariant, i.e. 
\begin{equation}\label{rotinvcond}
p_{1, a + \delta}(s_{1} + \delta) = p_{1, a}(s_{1}) \quad ; \quad 
p_{2, b + \delta}(s_{2} + \delta) = p_{2, b}(s_{2}) \qquad ; \qquad 
\forall \delta\in \Real
\end{equation}
and twice continuously differentiable, then the probability measure 
$dR_{a, b}(s_{1}, s_{2})$, defined by (\ref{ChamMeas}), must have either the form
\begin{equation}\label{rhoModulusCos}
dR_{a, b}(s_{1}, s_{2}) =  \delta(s_{1} - s_{2}) ds_{1} ds_{2} \frac{1}{4}|\cos(s_{1} -a)| 
\end{equation}
or the form
\begin{equation}\label{rhoModulusCos2}
dR_{a, b}(s_{1}, s_{2}) =  \delta(s_{1} - s_{2}) ds_{1} ds_{2} \frac{1}{4}|\cos(s_{2} - b)|.
\end{equation}
\end{theorem}
\Proof
Because of rotation invariance
\begin{eqnarray*}
p_{1, a}(s_{1}) &=& p_{1, 0}(s_{1} -a) =:p_{1}(s_{1} - a)\\
p_{2, b}(s_{2}) &=& p_{2, 0}(s_{2} -b) =:p_{2}(s_{2} - b).
\end{eqnarray*}
Using the result of Proposition (\ref{DiagonalSupportZero}), we have
\[
dR_{a,b}(s_1,s_2) = \rho(s_1) p_{1}(s_1 - a)p_{2}(s_2 -b) \delta(s_1-s_2)ds_1ds_2.
\]
For $a$ and $b$ satisfying $0 \leq b -a \leq \pi$, $I_{a} \cap I_{b} = [-\pi/2 + b, a + \pi/2)$, and therefore
\[
R_{a, b}({I_{a}} \times {I_{b}})
=
\int_{-\pi/2 + b}^{a + \pi/2} ds_{1} \rho(s_{1}) p_{1}(s_{1} - a) p_{2}(s_{1} - b).
\]
By (\ref{QMP}),
\[
\frac{1}{4}( 1 + \cos(b-a) )= R_{a, b}(I_{a}\times I_{b})=
\int_{-\pi/2 + b}^{a + \pi/2} ds_{1} \rho(s_{1}) p_{1}(s_{1} - a) p_{2}(s_{1} - b).
\]
Differentiating this with respect to $b$, we have
\begin{eqnarray}
-\frac{1}{4}\sin(b -a) 
&=& - \rho(b - \pi/2)p_{1}(b- a -\pi/2) p_{2}(-\pi/2)  \nonumber \\
&& + \int_{b -\pi/2}^{a + \pi/2} ds_{1} \rho(s_{1}) p_{1}(s_{1} - a) p_{2}'(s_{1} - b).
\label{FirstDeriv}
\end{eqnarray}
Putting $b= a + \pi$, we obtain
\[
0 = \rho(a + \pi/2) p_{1}(\pi/2)  p_{2}(-\pi/2).
\]
Since $a$ is arbitrary and $\rho$ is a probability density, $\rho(a + \pi/2)$ cannot vanish.
Hence
\[
p_{1}(\pi/2) = 0 \ \mbox{ or } \ p_{2}(-\pi/2)=0.
\]

Let us assume that $p_{1}(\pi/2) = 0$.
Differentiating (\ref{FirstDeriv}) with respect to $b$ and putting $b= a+ \pi$, we obtain
\[
\frac{1}{4}
=
-\rho(a + \pi/2) p_{1}'(\pi/2) p_{2}(-\pi/2).
\]
From this we can see
\[
p_{1}'(\pi/2)\not=0 \ \mbox{ and } \ p_{2}(-\pi/2)\not= 0
\]
and $\rho(a + \pi/2) = 1/(4 p_{1}'(\pi/2) p_{2}(-\pi/2))=$ const.,
since $a$ is arbitrary.
Thus we write $\rho(s_{1}) = c$ hereinafter.

Since $I_{a} \cap J_{b} = [-\pi/2 + a, -\pi/2 + b)$,
by (\ref{QMP})
\[
\frac{1}{4}( 1 - \cos(b-a) )= R_{a, b}(I_{a}\times J_{b})=
c \int_{-\pi/2 + b}^{-\pi/2 + a} ds_{1}  p_{1}(s_{1} - a) p_{2}(s_{1} - b).
\]
Differentiating this with respect to $b$, we have
\begin{eqnarray*}
\frac{1}{4} \sin (b - a)
&=& -c p_{1}(b- a -\pi/2) p_{2}(-\pi/2) \\
&&+ c \int_{-\pi/2 + b}^{-\pi/2 + a} ds_{1}  p_{1}(s_{1} - a) p_{2}'(s_{1} - b).
\end{eqnarray*}
Putting $b = a$, we obtain
\[
0 = - c p_{1}(-\pi/2)  p_{2}(-\pi/2). 
\]
Since $p_{2}(-\pi/2)\not=0$,
\[
p_{1}(-\pi/2) = 0.
\]

Since $(I_{a}\cap J_{b})\cup(I_{a}\cap I_{b}) = [-\pi/2 + a,\ a + \pi/2)$, by (\ref{QMP}) we have
\[
\frac{1}{2}= R_{a, b}(I_{a}\times J_{b} \cup I_{a}\times I_{b}) =
c \int_{-\pi/2 + a}^{a + \pi/2} ds_{1}  p_{1}(s_{1} - a) p_{2}(s_{1} - b).
\]
Changing variable with $s = s_{1} -a$, we obtain
\[
\frac{1}{2} = c \int_{-\pi/2 }^{\pi/2} ds  p_{1}(s) p_{2}(s - b + a).
\]
In the same way, for $\pi \leq b - a \leq 2\pi$, $I_{a} \cap I_{b} = [- \pi/2 + a, -3\pi/2 + b)$ and $I_{a}\cap J_{b} = [-3\pi/2 + b, a + \pi/2)$, we have
\begin{eqnarray*}
\frac{1}{2} &=& 
R_{a, b}(I_{a}\times I_{b} \cup I_{a}\times J_{b}) =
c \int_{-\pi/2 + a}^{a + \pi/2} ds_{1}  p_{1}(s_{1} - a) p_{2}(s_{1} - b) \\
&=&
c \int_{-\pi/2}^{\pi/2} ds  p_{1}(s) p_{2}(s - b + a).
\end{eqnarray*}
Since $p_{1}$ is continuous and $p_{1}(\pi/2) = p_{1}(-\pi/2)=0$ and $a$ and $b$ are arbitrary, we can see that $ p_{2}(s)=$const.$=: c_{2}$.
Thus by renaming
\[
\widetilde{p}_{1}(s_{1}) := c p_{1}(s_{1}) c_{2} 
\]
we find
\[
dR_{a,b}(s_1, s_2) = \widetilde{p}_{1}(s_1 - a) \delta(s_1 - s_2)ds_1ds_2.
\]

Our remaining task is to determine the form of $\widetilde{p}_{1}$.
For $a$ and $b$ satisfying $0 \leq b -a \leq \pi$, (\ref{FirstDeriv}) becomes
\[
-\frac{1}{4}\sin(b-a) = - \widetilde{p}_{1}(-\pi/2 + b - a).
\]
By putting $s = b -\pi/2$, $\widetilde{p}_{1}(s - a) = \frac{1}{4} \cos(s-a)$ 
for $-\pi/2 \leq s - a \leq \pi/2$.
Therefore 
\[
\widetilde{p}_{1}(s -a) = \frac{1}{4} |\cos(s-a)|,\  -\pi/2 \leq s - a \leq \pi/2.
\]

Since $J_{a} \cap I_{b} = [a + \pi/2, b + \pi/2)$,
\[
\frac{1}{4}( 1 - \cos(b-a) )=  R_{a, b}({J_{a}} \times {I_{b}})
=
\int_{a + \pi/2}^{b + \pi/2} ds_{1} \widetilde{p}_{1}(s_{1} - a).
\]
By differentiating this with respect to $b$ we have
\[
\frac{1}{4}\sin(b-a) =  \widetilde{p}_{1}(b + \pi/2 - a).
\]
By putting $s = b + \pi/2$, $\widetilde{p}_{1}(s - a) = 
\frac{1}{4} \sin(s-a -\pi/2) = -\frac{1}{4} \cos(s-a)$ 
for $\pi/2 \leq s - a \leq 3\pi/2$.
Therefore 
\[
\widetilde{p}_{1}(s - a) = \frac{1}{4} |\cos(s-a)|,\  \pi/2 \leq s - a \leq 3\pi/2.
\]

Accordingly, 
\[
dR_{a, b}(s_{1}, s_{2}) = \delta(s_{1} - s_{2}) ds_{1} ds_{2} \frac{1}{4}|\cos(s_{1} - a)|.
\]

If we assume that $p_{2}(-\pi/2)=0$ instead of $p_{1}(\pi/2) = 0$, then in the same way we obtain
\[
dR_{a, b}(s_{1}, s_{2}) = \delta(s_{1} - s_{2}) ds_{1} ds_{2} \frac{1}{4}|\cos(s_{2} - b)|.
\]
\QED

\section{Two experimental settings for determinism}\label{2setfordet}

In classical statistical mechanics the dynamical evolution is deterministic but 
the initial information is incomplete and is represented by a probability measure
which describes the preparation of the experiment.

In the case of adaptive systems however the experimental setup is not fully 
determined at the initial time in the sense that many measurements are a priori possible 
and the particles don't know which one will be actually performed. This means that part of
the dies are cast at the source, where the particles are emitted, and part of
the dies are cast at the final time, when each particle interacts with the measurement apparatus.

It is clear that the two experimental situations must correspond to different mathematical models. In the present section we try to make these differences explicit.

Standard determinism can be summed up in the statement: the state at any time $t=t_0$ uniquely determines the states at any later time $(t>t_0)$. For reversible determinism also the converse is true: the state at any time $T$ uniquely determines the state at any time $t_0<T$. In exact deterministic theories states are characterized by the values of some observables, like position and momentum in classical mechanics.

We call ``configuration (or phase) space'' the state space of an exact deterministic theory.

In statistical deterministic theories, one postulates the existence of an underlying exact theory and the states are probability measures on the configuration space of this theory. The prototype example is classical statistical mechanics and the models considered in the present paper fall into this category, i.e. a statistical, reversible deterministic theory.

The mathematical model of such a theory is defined by
\begin{itemize}
\item[--] a configuration space $\Omega$

\item[--] a deterministic, reversible dynamics $T^t:\Omega\to\Omega$

\item[--] a probability measure $P$ on $\Omega$.
\end{itemize}
The interpretation of $P$ depends on the experimental setting. We distinguish two cases:
\begin{itemize}
\item[(i)] $P$ condensates the experimental information available at an initial time $t_0$

\item[(ii)] $P$ condensates the experimental information available at a final time $t_f$, i.e. the time when the experiment is actually performed.
\end{itemize}

According to von Neumann measurement theory a mathematical description of a measurement process must take into account the interaction of the measured system with the measurement apparatus.

This means that, for adaptive systems (like chameleons) the meaning of the probability measure $P$ must be understood in the sense of (ii) above.

More precisely, von Neumann measurement scheme requires the specification of:
\begin{itemize}
\item[--] 
 a configuration space $M$ of the apparatus 
\item[--]
 a joint dynamics
$$T^t_{S,M}:S\times M\to S\times M$$
describing the evolution of the composite system (system, apparatus).
\end{itemize}
In the case of adaptive systems, at the initial time $t_0$ one has a whole family of possible measurements and the one which will be performed will be known only at the final time $t_f$.

Therefore a von Neumann type description of an adaptive system should consist of a multiplicity of triples
$$(S\times M, T^t_{S,M},P_{S,M})$$
i.e. on triple for each of the possible measurements.

Moreover, since the choice of the measurement, and therefore all the available experimental data, occur at a final time $t_f$, the identity
$$P_{t_0}=T^{-(t_f-t_0)}_{S,M}P_{S,M}$$
which expresses the unknown initial distribution $(P_{t_0})$ in terms of the experimentally found distribution $(P_{S,M})$, shows that the initial distribution depends on the measurement. This circumstance does not violate the causality principle because such an initial distribution should be interpreted as the conditional distribution at time $t_0$ of the composite system $(S,M)$ given the knowledge of the results of the experiment $M$, performed at time $t_f>t_0$.

The local causal measures discussed in the present paper correspond to the final measures $P_{S,M}$ described here.


\section{Empirical correlations of systems of distant particles}\label{empcordistpa}

In the present section we argue that the same term ``pair correlation'' is used 
to describe two completely different experimental procedures and that a good 
mathematical model should take into account these experimental differences.

If $S$ is the configuration space of a classical system, then by definition a 
trajectory of this system is a map
\[
\sigma:t\in[t_\sigma,+\infty)\mapsto\sigma_t\in S.
\]
For each $t\in[t_\sigma,+\infty)$, $\sigma_t$ is interpreted as the configuration of the system at time $t$. 
In the following we fix the interval $[t_\sigma,+\infty)$ and we often will not mention it.

If $(1,2)$ denotes a composite system made of two particles, a trajectory of the pair is by definition a pair $(\sigma_1,\sigma_2)$, where $\sigma_1$ is a trajectory of particle 1 and $\sigma_2$ is a trajectory of particle 2.

We suppose that all the particles $1_j$ (resp. $2_j$), $j \in \{ 1, \ldots , N\}, N\in \Ntr$, have the same configuration space $S_1$ (resp. $S_2$) so that all the $\sigma_{1,j}$ (resp. $\sigma_{2,j}$) are functions
\[
\sigma_{1,j}:[t_{\sigma_1}, +\infty)\to S_1\quad(\hbox{resp. }\sigma_{2,j}:[t_{\sigma_2},+\infty)\to
S_2).
\]

Let $(f_1,f_2)$ be an observable of the pairs $(1_j,2_j)$.
The term {\it empirical correlation between $f_1$ and $f_2$\/} has a multiplicity of meanings depending on the experimental procedure employed to measure this quantity.
In the following we shall describe these possibilities which are frequently met.

By definition of classical system, if a configuration space of a system is $S$, an observable of the system is a real valued function $f$ defined on $S$, i.e., 
$f:S\to\Real$.
An observable of a pair of systems $(1,2)$ is a pair $(f_1, f_2)$, where $f_1$ is an observable of system 1 and $f_2$ is an observable of system 2.

If it is given an ensemble of pairs
\begin{equation}
(1_j,2_j),\ \quad j\in\{1,\dots,N\},  \label{ensps}
\end{equation}
$(\sigma_{1,j},\sigma_{2,j})$
denotes the trajectory of the $j$th pair ($j=1,\ldots,N$).
If this ensemble of pairs is obtained by repeating measurements with the same measurement apparata on successively emitted particles from a source, then $t_{\sigma_{1, 1}} < \cdots < t_{\sigma_{1, N}}$, 
 $t_{\sigma_{2, 1}} < \cdots < t_{\sigma_{2, N}}$.

To fix the ideas, from now on we shall think of a source which emits pairs of particles and particles of a pair are emitted simultaneously, i.e.,
\[
t_{\sigma_{1,j}}=t_{\sigma_{2,j}}=t_j
\]
for each trajectory  $(\sigma_{1j},\sigma_{2j})$ in concrete experimental situations.

\subsection{Standard correlations}

The term {\it standard correlation} is used when the following physical conditions are verified:
\begin{enumerate}
\def\labelenumi{\arabic{enumi})}
\item 
The total number $N$ of pairs is exactly known.

\item 
The trajectory of each pair can be followed without disturbance so that, at each time $t$, the experimenters know exactly to which of the pairs (\ref{ensps}) their measurement is referred. This property will be called {\it distinguishability}.

\item 
The observable $(f_1,f_2)$ is measured on each pair of the ensemble. The result of the measurement of $(f_1,f_2)$ on the $j$th pair will be denoted by
\[
(f_{1,j},f_{2,j});
\]
the measurement itself will be denoted by $M_j$.
\end{enumerate}
Under these conditions the following definition makes sense.
\begin{definition}\label{stempcodf}
The empirical correlation between the pair of observables $(f_1,f_2)$, relative to the sequence of measurements $M=(M_j)$ on the ensemble $\{(1_j,2_j):\ j= 1,\ldots, N \}$ is
\begin{equation}
\langle f_1\cdot f_2\rangle_M:={1\over N}\,\sum^N_{j=1}f_{1,j}f_{2,j}.
\label{dfempco}
\end{equation}
\end{definition}
We further specify our context of standard correlations as follows.
\begin{enumerate}
\setcounter{enumi}{3}
\def\labelenumi{\arabic{enumi})}
\item 
Each measurement $M_j$ is specified by a time
\[
t'_j:=t_j+T,
\]
where $T$ is independent of $j$ (recall that $t_j$ is the emission time for the pair $(1_j,2_j)$).

\item 
The result of the $j$th measurement does not depend on the interval $[t_j,t_j+T]$ but only on $T$ (time homogeneity).
\end{enumerate}
Under these conditions the correlations (\ref{dfempco}) are interpreted as the correlations of $(f_1,f_2)$ at time $T$ and $T$ is interpreted as the final time of the single measurement.

\subsection{Correlations of distant pairs}

Suppose that the measurement protocol is the following.
\begin{itemize}
\item[(DP1)] 
It is known that each pair is emitted simultaneously, but the experimenters do not know precisely when, i.e., $t_{\sigma,j}$ is not known.
\item[(DP2)] 
The experimenters cannot follow the trajectory of each particle, but only register the  result of a measurement at time $t$ ({\it indistinguishability}).
\item[(DP3)] 
The experimenters have synchronized clocks, so the time $t$ is the same for both.
\item[(DP4)] 
The experimenters do not know the total number of emitted particles.
\item[(DP5)] 
The experimenters cannot postulate that, if a particle of a pair reaches one of them, then the other particle reaches the other experimenters.
\end{itemize}

Conditions (4) and (5) of the previous section are still meaningful because they are referred to single particles. However condition (3) is meaningless because of indistinguishability.
Moreover the $N$, in formula (\ref{dfempco}) is unknown. In a situation described by the above conditions we speak of {\it correlations of distant particles\/}.

In conclusion: under the above described physical conditions, the definition of standard correlations is meaningless and a new one is needed.

\begin{definition}\label{codipadf} 
The protocol to define correlations of distant particles is the following:
\begin{itemize}
\item[(CDP1)] 
The experimenter $X$, $X\in\{1,2\}$ performs measurements on $M_X$ particles and records
\begin{itemize}
\item[--] the time $t'_{X,j}$ of the $j$th measurement
\item[--] the value $f_{X,j}$ of the measured observable $f_X$
\end{itemize}
for $\forall\,j\in\{1,\dots,M_X\}$.

\item[(CDP2)] 
The two experimenters exchange the sequences
\[
\Bigl( (t'_{1,j},f_{1,j}): \  j=1,\dots,M_{1} \Bigr)
\mbox{ and }
\Bigl( (t'_{2,j},f_{2,j}): \  j=1,\dots,M_{2} \Bigr).
\]
\item[(CDP3)] 
Each experimenter extracts the sequences
\[
\Bigl(f'_{1,h} :\  h = 1,\dots,M_{f_1f_2} \Bigr)
\mbox{ and }
\Bigl(f'_{2,h} :\  h = 1,\dots,M_{f_1f_2} \Bigr),
\]
where
\[
\Bigl\{ s_h :\  h \in \{ 1,\dots,M_{f_1f_2}\} \Bigr\}
:=
\Bigl\{ t'_{1,j}:j\in\{1,\dots,M_1\} \Bigr\} 
\cap
\Bigl\{ t'_{2,j}:j\in\{1,\dots,M_2\} \Bigr\}
\]
and
\[
f'_{X, h} := f_{X, j}, \mbox{ if } s_{h} = t'_{X, j}\qquad (X=1, 2).
\]
\item[(CDP4)] 
The empirical correlations of distant pairs are defined by
\[
\langle f_1f_2\rangle_{DP}:={1\over M_{f_1,f_2}}\,\sum^{M_{f_1,f_2}}_{h=1}
f'_{1,h}
f'_{2,h}.
\]
\end{itemize}
\end{definition}
In other words: by definition, correlation of distant pairs means conditioned 
correlations on coincidences.

\bigskip

\Remark
 Practically the totality of the EPR type experiments follow the protocol described 
in Definition \ref{codipadf}.

\subsection{Mathematical models of empirical correlations}

We keep the notations introduced in the previous sections. 
Instead of considering a single observable for each particle of a pair, 
we consider now two families of observables:
$\hat{\cal A}_1$ -- of particles of type 1, $\hat{\cal A}_2$ -- of particles of type 2.
We suppose that, for each pair
$$
\hat S_{1,a}\in\hat{\cal A}_1\ ;\quad\hat S_{2,b}\in\hat{\cal A}_2
$$
one has performed experiments leading to estimates of all the empirical correlations 
\[
\kappa_{a,b}:=\langle\hat S_{1,a}\hat S_{2,b}\rangle_{EMP}
\]
These numbers are experimental data.

We suppose moreover that the experimental protocols to determine these correlations have 
been homogeneous, e.g., always standard correlations or always distant pair correlations.

\begin{definition} \label{MMEC}
A mathematical model for the empirical correlations $\{\kappa_{ab}\}$ is defined by:
\begin{itemize}
\item[--] a family of probability spaces $(\Omega,{\cal F},P_{a,b})$ where the pairs $(a,b)$
label the a priori possible experimental settings

\item[--] two families ${\cal A}_1$, ${\cal A}_2$ of real valued functions on 
$\Omega$ with the property that 
$\forall S_{1,a}\in{\cal A}_1, \forall S_{2,b}\in{\cal A}_2$, one has
\begin{equation}\label{emp-cor-ab}
\kappa_{a,b}=\int_\Omega S_{1,a}S_{2,b}dP_{a,b}
\end{equation}
\end{itemize}
Such a model is called {\it local\/} if there exists a computer program which allows 
to simulate the protocol of the experiment in such a way that:
\begin{itemize}
\item[--] the program must run on three non-communicating computers: 
Computer $S$, Computer 1, Computer 2.

\item[--] Computer $S$ should produce a family of pairs $(\sigma_{1,j},\sigma_{2,j})$, $j\in\{1,\dots,N\}$ without using any information on what Computers 1 and 2 will do. 
Then Computer $S$ sends $(\sigma_{1,j})$ to Computer 1 and $(\sigma_{2,j})$ to Computer 2;

\item[--] Computer 1 (resp. 2) should for each $j\in\{1,\dots,M\}$
\begin{enumerate}
\def\labelenumi{(\roman{enumi})}
\item 
choose one observable 
\[
S_{1,a}\in{\cal A}_1 \ (\hbox{resp. }S_{2,b}\in{\cal A}_2),
\]

\item 
compute the configuration $\sigma_{1,j,a}(T)$ of particle $1_j$ at time $T$ using 
only informations on the trajectory $\sigma_{1,j}$ and the observable $S_{1,a}$ 
(resp. $ \sigma_{2,j,b}(T); \sigma_{2,j}, S_{2,a}$),

\item 
check if $\sigma_{1,j,a}(T)\in W$ where $W\subseteq S_1=S$ is window of the 
configuration space (resp. $\sigma_{2,j,b}(T)\in W$).

This simulates the physical phenomenon that certain local trajectories of the 
particles may end up outside the phase space window defining the coincidence.

\item 
In case $\sigma_{1,j,a}(T)\in W$ (resp. $\sigma_{2,j,b}(T)\in W$), compute the value
$S_{1,a}(\sigma_{1,j}(T))$ (resp. $S_{2,b}(\sigma_{2,j}(T))$.

\item 
The procedure to compute the correlations must reproduce exactly the procedure used in the corresponding experimental protocol and described by Definition (\ref{codipadf}).
\end{enumerate}
\end{itemize}
\end{definition}

The model in Ref. \cite{uchiyama95} can be considered as a local mathematical model for 
the empirical correlations in Definition \ref{MMEC}, if the protocol for distant pairs 
is adopted, although this model reproduces the EPR correlations only approximately.

\section*{Acknowledgment} 
The author (S. U.) thanks Professor A. Yu. Khrennikov for the useful discussion in the early stages of this work.
He thanks people of the Volterra Center for their warm hospitality and kindness during his long visit.
He also thanks his colleagues at Hokusei Gakuen University Junior College for their support.


\end{document}